\documentclass[article]{revtex4}
\usepackage{pdfpages}
\usepackage{epsf}
\usepackage{subfigure}
\usepackage{amsmath}
\usepackage{eqnarray,amsmath}
\usepackage{epstopdf}
\usepackage{mathrsfs}
\usepackage{natbib}
\usepackage{amssymb,latexsym}
\usepackage{dcolumn}
\usepackage{graphicx}
\usepackage{color}

\makeatletter
\newcommand*{\rom}[1]{\expandafter\@slowromancap\romannumeral #1@}
\makeatother
\def\be{\begin{equation}}
    \def\ee{\end{equation}}
\def\ba{\begin{eqnarray}}
    \def\ea{\end{eqnarray}}

\graphicspath{{images/}}
\begin{document}
    \title{\large \bf  Constructing an entangled state in Heisenberg picture for inflationary cosmology}
    
    \author{Abasalt Rostami}
    \affiliation{Department of Physics, Sharif University of Technology,
        Tehran, Iran }
    \affiliation{ School of Physics, Institute for Research in Fundamental Sciences (IPM), P. O. Box 19395-5531, Tehran, Iran }
    \email{aba-rostami@ipm.ir}
    
    \author{Javad T. Firouzjaee}
    \affiliation{ School of Astronomy, Institute for Research in Fundamental Sciences (IPM), P. O. Box 19395-5531, Tehran, Iran }
    \email{j.taghizadeh.f@ipm.ir}
    \begin{abstract}
        The effects of the entanglement of the inflaton field which initially entangled with those of another field on observables like power spectrum are known in the context of the Schr\"{o}dinger field theory. To clarify this effect in Heisenberg picture, there were some attempts to construct the initial entangled state by making use of an entangled transformation (like Bogoliubov transformation) between the Bunch-Davies vacuums and squeezed states. We study the role of the time-dependent entangled transformation in the Schr\"{o}dinger field theory. We derive the relation between two vacuum states which their mode functions are transformed by the entangled transformation in  Schr\"{o}dinger picture. We discuss that the time-dependency of the entanglement parameter is inevitable and only in the first order of the entanglement parameter perturbation this time-dependency vanishes. We study the entangled transformation in the Heisenberg picture in term of the entangled parameter which appears in the entangled state in  Schr\"{o}dinger picture.

    \end{abstract}
    %
    %
    \maketitle
    \tableofcontents
    
    \newpage
    
    \section{Introduction}
    The main experimental observation of the early universe, Cosmic Microwave Background (CMB), have become very accurate and have allowed us to deduce the details of the even earlier phase of the cosmos \cite{Ade:2015xua}. It is known that the temperature fluctuation in the CMB are originated from the early quantum fluctuation which becomes classical as the universe expand. Therefore, studying the quantum nature of the inflationary background can lead to better understanding of the observables like power spectrum and bispectrum which can have quantum correction.\\

An important aspect of the quantum physics is the entanglement which has been one of the fascinating features predicted by quantum mechanics since Einstein-Podolsky-Rosen (EPR) \cite{EPR}. Although more attention has been paid to how to make use of quantum entanglement of EPR pairs in quantum cryptography and quantum information \cite{review-en}, the entanglement entropy plays an important role in studying the black hole information problem and holography which we can count of the degrees of entanglement of a quantum system. In the cosmology, Maldacena \cite{Maldacena:2012xp} developed an explicit method to calculate the entanglement entropy in a quantum field theory in the Bunch-Davies vacuum of the de Sitter space and showed that there can be the quantum entanglement between two causally disconnected open charts in de Sitter space and in this way more ideas presented to detect these quantum effects \cite{En-Observational}. Along the way,  one can study the quantum entanglement in the  Schr\"{o}dinger field theory which the operators are time independent. The  Schr\"{o}dinger picture is the natural framework to study the entanglement between the fields \cite{Holman-entanglement}. The  Schr\"{o}dinger representation is also best suited to discuss the decoherence of the cosmological perturbations and the quantum to classical transition in inflationary models \cite{Polarski:1995jg}  and one can develop in-in formalism in the  Schr\"{o}dinger picture for curvature perturbations \cite{Rostami:2016ulp} to study N-point function in the observation. In the way, many works have been done to study the entanglement effect and the quantum to classical transition in the inflationary models \cite{Martin:2015qta}. 
On the other hand, apart from this fact that the Heisenberg picture field theory is more common and easier to study the inflationary models (for example the calculation of the Power spectrum and non-gaussianity in Schr\"{o}dinger picture needs the heavy calculation on the functional space), this picture has better description for particle representation of each state with explicit presentation of annihilation and creation process. Moreover, constructing the entangled states using the entanglement transformation for fermions is easier in the Heisenberg picture than Schr\"{o}dinger picture \cite{kanno-sasaki}.   There is an instruction to built the entangled state in Heisenberg picture.

Recently,  in \cite{Kanno:2015ewa} an initially entangled state between two free massive scalar fields in de Sitter space was studied and the 
 entangled state using a Bogoliubov transformation between the Bunch-Davies vacuum and a four-mode squeezed state was constructed. In this paper, we show that though Kanno \cite{Kanno:2015ewa} presented a novel way to study the entangled state in the Heisenberg picture, her calculation and analysis to introduce the entanglement was not correct in some parts. We follow her way to study the role of the time-dependent entangled transformation in the Schr\"{o}dinger field theory. We derive the relation between two vacuum states which their mode functions are transformed by the entangled transformation and also Bogoliubov transformation in  Schr\"{o}dinger picture. We calculate the entangled parameter in  Schr\"{o}dinger picture in term of the time-dependent entangled transformation in the Heisenberg picture, and the inflation power spectrum will be calculated in term of the time-dependent entangled parameter in Heisenberg picture. It is discussed that the time-dependency of the entanglement parameter is inevitable and only in the first order of the entanglement parameter perturbation this time-dependency vanishes. We will discuss the limitation that the current definition of the entanglement.\\

The paper is organized as follows. First, we review the quantization of a mono-field for Friedmann-Lemaitre-Robertson-Walker (FLRW) background in Section II. Section III is devoted to introduce and use the Bogoliubov transformation to get the vacuum state in the Schr\"{o}dinger picture. In Section IV we introduce the ABH entangled state in the Schr\"{o}dinger picture of the inflationary cosmology. Next, after discussing the relation between the ABH entangled state and the entangled stated which built of the entangled transformation, the inflationary power spectrum will be calculated in Section V. We end with a discussion in Section VI.
\\

\section{A Short Review On Mono-field Quantization }
The main goal of this section is to study a free scalar field in time-dependent gravitational backgrounds. At first step, we shall consider a free scalar field in a homogeneous and isotropic universe and then try to quantize it. In this way, we follow Mukhanov and Winitzki book notation \cite{mukhanov-book}.

A minimally coupled real scalar field φ(x) in a curved spacetime is described by the action
\begin{equation}\label{eq1}
S[\phi]= \int d^4 x \sqrt{-g} \left( \frac{1}{2} g^{\alpha\beta} \partial_{\alpha} \phi \partial_{\beta} \phi - \frac{1}{2} m^2 \phi^2 \right).
\end{equation}

Now, we consider an important class of metrics which is used for the homogeneous and isotropic  space-time namely FLRW. This is  characterized by the following metric
\begin{equation}\label{eq2}
ds^2 = - dt^2 +  a^2 (t) d\vec{x}^2.
\end{equation}
where $a(t)$ is the scale factor. A flat FRW space-time is a conformally flat space-time. If one replaces the coordinate $t$ by the conformal time $\eta$,  the metric would be transformed  into a conformally flat form
\begin{equation}\label{eq3}
\eta (t) \equiv \int\limits_{{t_0}}^t {\frac{{dt}}{{a(t)}}}.
\end{equation} 
By this transformation the action (\ref{eq1})  is written as
\begin{equation}\label{eq4}
S[\phi]=  \frac{1}{2} \int d^3 x d\eta\  a^2 \left({ \phi^{\prime}}^{2} - (\nabla \phi)^2 - m^2 a^2 \phi^2 \right).
\end{equation}
where a prime denotes
the derivative with respect to the conformal time $\eta$.  This action gives the following equation of motion for our scalar field
\begin{equation}\label{eq5}
 {\phi^{''}}^{2} - \nabla^2 \phi +2 \frac{a^{'}}{a} \phi^{'} + m^2 a^2 \phi = 0.
\end{equation}
Although this equation seems a bit different from the equation of motion which one derives for a scalar field in flat space-time but by using of the field redefinition $\tilde{\phi} = a \phi$, we could make it more similar to the equation of motion a scalar field in flat background. In fact with this redefinition the action for $\tilde{\phi}$ is
\begin{equation}\label{eq6}
S[\tilde{\phi}]=  \frac{1}{2} \int d^3 x d\eta  \left({\tilde{\phi^{'}}}^{2} - (\nabla \tilde{\phi})^2 - m^2 (\eta)\  \tilde{\phi}^2 \right)
\end{equation}
where, we have defined
\begin{equation}\label{eq7}
m^2 (\eta)= m^2 a^2 - \frac{a^{''}}{a}.
\end{equation}
Now, the equation of motion for the new field is
\begin{equation}\label{eq8}
 {\tilde{\phi^{''}}} - \nabla^2 \tilde{\phi} + m^2 (\eta) \tilde{\phi}  = 0
\end{equation}
this equation is like what we had in flat space-time for a free field with this difference that here the mass term is time dependent. All information about the influence of the gravitational field on $\tilde{\phi}$ is encapsulated in the time-dependent mass $m(\eta)$. This would lead  to this fact that the energy of the field $m(\eta)$ is generally not conserved. In quantum physics language, this leads to the possibility of particle creation which the energy for new particles is supplied by the gravitational field.
If we expand the field $\tilde{\phi}$ in Fourier modes, we have
\begin{equation}\label{eq9}
\tilde{\phi} (\eta, x) = \int \frac{d^3 k}{(2\pi)^3} \tilde{\phi}_{k} (\eta) e^{i k.x}.
\end{equation}
Thus, reach to the the decoupled equations of motion for the modes $\tilde{\phi}_k$
\begin{equation}\label{eq10}
 {\tilde{\phi^{''}}}_{k}^{2} + \omega_{k}^2 (\eta) \tilde{\phi}_{k}  = 0, 
 \end{equation}
 where we have defined
 \begin{equation}\label{eq11}
\omega_{k}^2 (\eta) \equiv k^2 + m^2 (\eta).
\end{equation}
This equation has two linearly independent solutions for each mode which these two solutions span the two-dimensional space of all solutions. If $v_k (\eta)$ is a complex solution then $v_{k}^{*} (\eta)$ would be another independent solution.  Any two linearly independent solutions have to have a non-vanishing Wronskian. One can easily show the Wronskian of any two independent solutions of (\ref{eq10}) is constant in conformal time. We choose this constant  as following
 \begin{equation}\label{eq12}
W(v_k, v^{*}_{k})=v_{k}^{'}v^{*}_{k} - v_{k} {v^{'}}^{*}_{k}= 2i.
\end{equation}
The general solution $\tilde{\phi}_k$ can be expressed as a linear combination of $v_{k}^{*} (\eta)$ and $v_{k}(\eta)$ as
\begin{equation}\label{eq13}
\tilde{\phi}_{k} (\eta) = \frac{1}{\sqrt{2}}[ a_k v_{k}^{*} (\eta) + a^{*}_{-k} v_{k}(\eta)],
\end{equation}
where the  $v_{k}^{*} (\eta)$ and $v_{k}(\eta)$ are called mode functions. Now, by replacing (\ref{eq13}) in (\ref{eq9}), we obtain $\tilde{\phi}(\eta, x)$ 
\begin{equation}\label{eq14}
\tilde \phi (\eta ,x) = \int {\frac{{{d^3}k}}{{{{(2\pi )}^{\frac{3}{2}}}}}\frac{1}{{\sqrt 2 }}\left[ {{a_k}v_k^*(\eta ){e^{ik.x}} + a_k^*{v_k}(\eta ){e^{ - ik.x}}} \right]} .
\end{equation}
Note that, the reality condition for $\phi (\eta ,x)$ entails $\tilde{\phi}_k= \tilde{\phi}^{*}_{-k}$.
To obtain a quantum picture of such theory, one has to consider $a_k$ and its conjugate $a_{k}^{\dagger}$  as a operator in given Hilbert space so at first we should write down $a_k$ as a combination of  $\tilde{\phi}_k$ and its conjugate momentum. Using the equation (\ref{eq13}) and (\ref{eq12}) we find $a_k$
\begin{equation}\label{eq15}
a_k = \frac{1}{i\sqrt{2}} ( v_{k}^{'} (\eta) \tilde{\phi}_{k} (\eta) - \tilde{\phi}_{k}^{'} (\eta) v_{k}(\eta) ),
\end{equation}
and its conjugate
\begin{equation}\label{eq16}
a_k^{*} = -\frac{1}{i\sqrt{2}} ({ v_{k}^{'}}^{*} (\eta) \tilde{\phi}_{-k} (\eta) - \tilde{\phi}_{-k}^{'} (\eta) v_{k}^{*}(\eta) ).
\end{equation}
Since the conjugate momentum of $\tilde{\phi}$ is
\begin{equation}\label{eq17}
\tilde{\Pi} (\eta, x) =\tilde{\phi}^{'} (\eta, x),
\end{equation}
 we have
\begin{equation}\label{eq18}
a_k = \frac{1}{i\sqrt{2}} ( v_{k}^{'} (\eta) \tilde{\phi}_{k} (\eta) - \tilde{\Pi}_{k} (\eta) v_{k}(\eta) )
\end{equation}
and
\begin{equation}\label{eq19}
a_k^{*} = -\frac{1}{i\sqrt{2}} ({ v_{k}^{'}}^{*} (\eta) \tilde{\phi}_{-k} (\eta) - \tilde{\Pi}_{-k} (\eta) v_{k}^{*}(\eta) ).
\end{equation}
Now, to reach a quantum picture, it is enough to define the usual following commutator
\begin{equation}\label{eq20}
\left[ \tilde{\phi} (\eta, x), \tilde{\Pi} (\eta, y) \right] = i \delta (x-y).
\end{equation}
This gives a representation for momentum as
\begin{equation}\label{eq21}
 \tilde{\Pi}_k (\eta) = -i (2\pi)^3 \frac{\delta}{\delta \tilde{\phi}_{-k} (\eta) }.
\end{equation}
The above equation leads to the usual commutation relation for creation and annihilation operators
\begin{equation}\label{eq22}
\left[ a_{k}, a^{\dagger}_{p} \right] = (2\pi)^3 \delta(k-p).
\end{equation}
Once the operators $a_k$ and $a_{k}^{\dagger}$ are determined, the vacuum state $\vert 0 \rangle$ is defined as the eigenstate of all annihilation operators $a_{k}$ with eigenvalue $0$, i.e. $a_{k} \vert 0 \rangle = 0$ for all k. An excited state $\vert n_{k_1}, m_{k_2}, \ldots \rangle$ with the occupation numbers $m, n, ...$ in the modes $\tilde{\phi}_{k_1}, \tilde{\phi}_{k_2}, \ldots $  is then given by
\begin{equation}\label{eq23}
\vert{n_{{k_1}}},{m_{{k_2}}}, \cdots \rangle  = \frac{1}{{\sqrt {n!m! \ldots } }}\left[ {{{({a_{{k_1}}}^\dag )}^n}{{({a_{{k_2}}}^\dag )}^m} \cdots } \right]\vert{0_{{k_1}}},{0_{{k_2}}}, \cdots \rangle
\end{equation}
It is interesting to know what is Schr\"{o}dinger picture of vacuum state. After determining this state one could use (\ref{eq19}) and (\ref{eq22}) to derive exited states in this picture. To derive the explicit form of vacuum, we use this fact that
\begin{equation}\label{eq24}
\langle\tilde{\phi_{k}}\vert a_{k} \vert 0\rangle = 0,
\end{equation}
where $\vert \tilde{\phi_{k}}\rangle$ is eigenstate of $\tilde{\phi_{k}}$.
Now using (\ref{eq18}), we will get a simple functional differential equation with the following solution 
\begin{equation}\label{eq25}
\Psi[\tilde{\phi}]=\langle\tilde{\phi_{k}} \vert 0\rangle = N(\eta) \exp\left[ \frac{i}{2} \int \frac{d^3 k}{(2\pi)^3} \frac{v^{'}_{k}}{v_k}(\eta)\ \tilde{\phi}_{k} \tilde{\phi}_{-k} \right]
\end{equation}
where $N(\eta)$ is the normalization factor and can be calculated by using of
\begin{equation}\label{eq26}
\int {D\tilde \phi }\  \Psi [\tilde \phi ]{\Psi ^*}[\tilde \phi ] = 1
\end{equation}
for the measure $D\tilde {\phi}=d\tilde{\phi}_{k_1} d\tilde{\phi}_{{-k}_1} d\tilde{\phi}_{k_2} d\tilde{\phi}_{{-k}_2} \ldots $. It is easy to show that apart of a constant for a given $k$-mode, we have $N_k \sim \frac{1}{|v_k (\eta)|}$.
 In the mathematical point of view, the wave function $\Psi[\tilde{\phi}]$ is defined on functional space of an infinite dimensional manifold. People rarely use this picture of quantum field theory because occupation number of each state is not manifest in this picture. However, some kind of calculations is easier than Heisenberg picture.
To end up this section  for future advantages, we derive wave function of the main field i.e. $\phi$. One can check out the annihilation and creation operators for this field would be 
\begin{equation}\label{eq27}
a_k = \frac{1}{i\sqrt{2}} \left[(a v_{k}^{'} (\eta)-a^{'} v_{k} (\eta)) \phi_{k} (\eta) +i(2\pi)^3 \frac{v_{k}}{a}(\eta) \frac{\delta}{\delta \phi_{-k}}  \right]
\end{equation}
and
\begin{equation}\label{eq28}
{a}^{\dagger}_k = -\frac{1}{i\sqrt{2}} \left[(a {v_{k}^{*}}^{'} (\eta)-a^{'} v_{k}^{*} (\eta)) \phi_{-k} (\eta) +i(2\pi)^3 \frac{{v_{k}}^{*}}{a}(\eta) \frac{\delta}{\delta \phi_{k}}  \right].
\end{equation}
Now, we could find the wave function as
\begin{equation}\label{eq29}
\Psi[\phi]=\langle\phi_{k} \vert 0\rangle = N(\eta) \exp\left[ \frac{i}{2} \int \frac{d^3 k}{(2\pi)^3}( \frac{v^{'}_{k}}{v_k}(\eta) - \frac{a^{'}}{a}(\eta))\ \phi_{k} \phi_{-k} \right]
\end{equation}
and we see that the difference is just $\frac{a^{'}}{a}$ in the exponent which called Hubble factor. Note that one could use the Schr\"{o}dinger equation in functional space to reach this wave function. We shall do it for two entangled scalar field later on.
\\

\section{Bogoliubov transformation}

Consider two sets of isotropic mode functions $u_k (\eta)$ and $v_k (\eta)$ are chosen. Since $v_k$ and $u_{k}^{*}$ are a basis, the function $u_k$ must be a linear combination of $v_k$ and $u_{k}^{*}$,
\begin{equation}\label{eq30}
u_{k} (\eta) = {\alpha}_k v_{k} (\eta) - {\beta}^{*}_{k} v_{k}^{*}(\eta)
\end{equation}
with time-independent complex coefficients $\alpha_k$ and $\beta_k$. If both sets $v_k$ and $u_k$ are
normalized by (\ref{eq12}), it follows that the coefficients $\alpha_k$ and $\beta_k$ satisfy
\begin{equation}\label{eq31}
|{\alpha}_k|^2  - |{\beta}_{k}|^2 = 1.
\end{equation}
Using the mode functions $u_k$ instead of $v_k$, one obtains an alternative mode
 expansion which defines another set $\tilde{a}_{k}$ and $\tilde{a}_{k}^{\dagger}$ of creation and annihilation operators
 \begin{equation}\label{eq32}
\tilde \phi (\eta ,x) = \int {\frac{{{d^3}k}}{{{{(2\pi )}^{\frac{3}{2}}}}}\frac{1}{{\sqrt 2 }}\left[ {{\tilde{a}_k}u_k^*(\eta ){e^{ik.x}} + \tilde{a}_k^*{u_k}(\eta ){e^{ - ik.x}}} \right]} ,
\end{equation}
in which we have defined
\begin{equation}\label{eq33}
\tilde{a}_{k} = {\alpha}_k a_{k} + {\beta}^{*}_{k} a_{-k}^{*}.
\end{equation}
 Now one can refer a new vacuum state $\vert \tilde{0}\rangle$ respect to $\tilde{a}_{k}$ which annihilated by this operator and then by expanding the new vacuum in term of previous basis (\ref{eq23}) to reach 
 \begin{equation}\label{eq34}
\vert \tilde{0}\rangle = \Pi_{k} \frac{1}{\sqrt{|\alpha_k|}} \exp \left( -\frac{\lambda_k}{2} a_{k}^{\dagger} a^{\dagger}_{-k} \right) \ \vert 0 \rangle
\end{equation}
where
\begin{equation}\label{eq35}
\lambda_k = \frac{\beta^{*}_k}{\alpha_k}.
\end{equation}
This is an infinite series which can be express explicitly for each mode as
\begin{equation}\label{eq36}
\vert{0_k},{0_{ - k}}\rangle  = \frac{1}{{|{\alpha _k}|}}\sum\limits_{n = 0}^\infty  {{{\left( { - \frac{{\beta _k^*}}{{{\alpha _k}}}} \right)}^n}} \vert{n_k},{n_{ - k}}\rangle.
\end{equation}
The $\alpha_k$ and $\beta_k$ constants are called Bogoliubov constants. It is interesting to know the closed exact form of the Bogoliubov transformation  in the Schr\"{o}dinger picture. To do it, we need to have the new vacuum in the Schr\"{o}dinger picture. It is obvious that
\begin{equation}\label{eq37}
\langle\tilde{\phi_{k}} \vert \tilde{0}\rangle =\tilde{N} \exp\left[ \frac{i}{2} \int \frac{d^3 k}{(2\pi)^3} \frac{u^{'}_{k}}{u_k}(\eta)\ \tilde{\phi}_{k}\tilde{\phi}_{-k} \right]
\end{equation}
because the difference between two vacuums is in mode functions and it is enough to replace $v_k$ with $u_k$. Using this fact that $|\lambda_k \frac{v^{*}_{k}}{v_k}| < 1$ and expanding $\frac{u_k^{'}}{u_k}$ in term of $\frac{v_k^{'}}{v_k}$, one obtains 
\begin{equation}\label{eq38}
\langle\tilde{\phi_{k}} \vert \tilde{0}\rangle = N \exp\left[ \frac{i}{2} \int \frac{d^3 k}{(2\pi)^3} \frac{v^{'}_{k}}{v_k}(\eta)\ \tilde{\phi}_{k}\tilde{\phi}_{-k} \right] \left(\prod_k \frac{1}{|\alpha_k|} \frac{1}{|1- \lambda_k \frac{v_{k}^{*}}{v_k}|}\right) \ exp\left(-\sum_{n=0}^{\infty} \int \frac{d^3 k}{(2\pi)^3} \lambda_{k}^{n+1} \frac{1}{v_{k}^2} (\frac{v_{k}^{*}}{v_k})^n \tilde{\phi}_{k}\tilde{\phi}_{-k} \right)
\end{equation}
This is a magic relation because not only expresses the new vacuum in a closed form of previous basis but also could be used to derive other exited states in the Schr\"{o}dinger picture. One can expand two last terms in the equation (\ref{eq38}) and compare $\lambda_{k}^{n}$ terms  with with their counterpart in (\ref{eq36}) to find all exited states.
Before ending this section, it is worth to mention a very important point that all information about kind of a state is hidden in mode functions. For example, by looking at (\ref{eq37}) one cannot observe any entanglement in this state. Therefore, to see this entanglement between excited states, we have to study mode functions of one observer and compare it with another observer.\\

\section{ABH State}

In 2014, Albrecht, Bolis and Holman (ABH) \cite{Holman-entanglement}  introduced an interesting state which describes entanglement between two non-interacting scalar fields in FLRW background. They applied it in Schr\"{o}dinger picture and  use it as an ansats to solve Schr\"{o}dinger equation. In this section, we have a short review of their derivation for more details see \citep{Holman-entanglement}. Suppose two fields $\phi$ and $\chi$ coupled to gravity with no any other interactions,
\begin{eqnarray}\label{eq39}
& & S= \frac{1}{2}\int d^4 x\ a^4(\eta)\left[\frac{1}{a^2(\eta)}\left(\Phi^{\prime}(\eta, \vec{x})^2-\left(\nabla \Phi(\eta, \vec{x})\right)^2\right) -m_{\Phi}^2\ \Phi(\eta, \vec{x})^2\right .+\nonumber\\
& & \left . \frac{1}{a^2(\eta)}\left(\chi^{\prime}(\eta, \vec{x})^2-\left(\nabla \chi(\eta, \vec{x})\right)^2\right) -m_{\chi}^2\ \chi(\eta, \vec{x})^2\right],
\end{eqnarray}
To be able to make a functional Schr\"{o}dinger equation for a such action one needs the Hamiltonian in terms of conjugate momenta;
\begin{eqnarray}\label{eq40}
& & H= \int d^3 x\ \left[\frac{\pi_{\Phi}^2}{2 a^2(\eta)}+\frac{1}{2} a^2(\eta) \left(\left(\nabla \Phi(\eta, \vec{x})\right)^2+ a^2(\eta) m_{\Phi}^2\ \Phi(\eta, \vec{x})^2\right)+\right .\nonumber\\
& & \left . \frac{\pi_{\chi}^2}{2 a^2(\eta)}+\frac{1}{2} a^2(\eta) \left(\left(\nabla \chi(\eta, \vec{x})\right)^2+ a^2(\eta) m_{\Phi}^2\ \chi(\eta, \vec{x})^2\right)\right],
\end{eqnarray}
where $\pi_{\phi}$ and $\pi_{\chi}$ are the canonically conjugate momenta for $\phi$ and $\chi$. Like before section, it would be more appropriate to go to Fourier space as following for $\phi$
\begin{equation}\label{eq41}
\Phi(\eta, \vec{x}) = \sum_{\vec{k}} \frac{\phi_{\vec{k}}}{\sqrt{V}} e^{-i \vec{k}\cdot \vec{x}},\quad \pi_{\Phi}(\eta, \vec{x}) = \sum_{\vec{k}} \frac{\pi_{\Phi, {\vec{k}}}}{\sqrt{V}} e^{-i \vec{k}\cdot \vec{x}},
\end{equation}
and with a similar transformation for $\chi$. It is easy to see with this decomposition, the Hamiltonian has the following form 
\begin{eqnarray}\label{eq42}
& &  H = H_{\Phi} + H_{\chi}\nonumber\\
& & H_{\Phi} = \sum_{\vec{k}} H_{\Phi, \vec{k}},\quad H_{\Phi, \vec{k}} = \frac{\pi_{\Phi, \vec{k}} \pi_{\Phi, -\vec{k}}}{2 a^2(\eta)} + \frac{1}{2} a^2(\eta)\left(k^2 + m_{\Phi}^2 a^2(\eta)\right) \phi_{\vec{k}} \phi_{-\vec{k}}\nonumber\\
& &  H_{\chi} = \sum_{\vec{k}} H_{\chi, \vec{k}},\quad H_{\chi, \vec{k}} = \frac{\pi_{\chi, \vec{k}} \pi_{\chi, -\vec{k}}}{2 a^2(\eta)} + \frac{1}{2} a^2(\eta)\left(k^2 + m_{\chi}^2 a^2(\eta)\right) \chi_{\vec{k}} \chi_{-\vec{k}}.
\end{eqnarray}
Now, in this level one can use the separation of variables method to solve Schr\"{o}dinger equation for each $k$-mode
\begin{equation}\label{eq43}
i\partial_{\eta} \psi_{\vec{k}} \left[\phi_{\vec{k}}, \chi_{\vec{k}}; \eta\right]=\left(H_{\Phi, \vec{k}}+H_{\chi, \vec{k}}\right)\psi_{\vec{k}} \left[\phi_{\vec{k}}, \chi_{\vec{k}}; \eta\right],
\end{equation}
 but such solution is only able to give an entangled state between exited states of one kind of fields not entanglement between different fields. To make a entangled state between different fields, we use the following ansats  
\begin{equation}\label{eq44}
\psi_{\vec{k}} \left[\phi_{\vec{k}}, \chi_{\vec{k}}; \eta\right] = N_k(\eta) \exp\left[-\frac{1}{2}\left(A_k(\eta) \phi_{\vec{k}} \phi_{-\vec{k}}+B_k(\eta) \chi_{\vec{k}} \chi_{-\vec{k}}+C_k(\eta)\left(\phi_{\vec{k}} \chi_{-\vec{k}}+\chi_{\vec{k}} \phi_{-\vec{k}}\right)\right)\right].
\end{equation}
where $C_k$ plays the role of entanglement between two fields. Inserting (\ref{eq43}) into (\ref{eq42}) and then matching the powers of the field modes gives us the following equations
\begin{eqnarray}\label{eq45}
& & i\frac{N_k^{\prime}}{N_k} = \frac{\left(A_k+B_k\right)}{2 a^2(\eta)} \nonumber\\
& & i A_k^{\prime} = \frac{A_k^2+C_k^2}{a^2(\eta)} -\Omega_{\Phi, k}^2 a^2(\eta),\quad \Omega_{\Phi, k}^2\equiv k^2 + m_{\Phi}^2 a^2(\eta)\nonumber\\
& &  i B_k^{\prime} = \frac{B_k^2+C_k^2}{a^2(\eta)} -\Omega_{\chi, k}^2 a^2(\eta),\quad \Omega_{\chi, k}^2\equiv k^2 + m_{\chi}^2 a^2(\eta)\nonumber\\
& &  i\frac{C_k^{\prime}}{C_k}=  \frac{\left(A_k+B_k\right)}{a^2(\eta)}.
\end{eqnarray}
One could reach a more suitable form of these equations  using the following definitions
\begin{equation}\label{eq46}
i A_k(\eta) \equiv a^2(\eta)\left(\frac{f_k^{\prime}(\eta)}{f_k(\eta)}- \frac{a^{\prime}(\eta)}{a(\eta)}\right),\quad i B_k(\eta) \equiv a^2(\eta)\left(\frac{g_k^{\prime}(\eta)}{g_k(\eta)}- \frac{a^{\prime}(\eta)}{a(\eta)}\right).
\end{equation}
Then the resulting differential equation under these definitions would be
\begin{eqnarray}\label{eq47}
& & f_k^{\prime \prime} + \left(\Omega_{\Phi, k}^2-\frac{a^{\prime \prime}(\eta)}{a(\eta)}\right) f_k = \frac{C_k(\eta)^2}{a^4(\eta)}f_k\nonumber\\
& & g_k^{\prime \prime} + \left(\Omega_{\chi, k}^2-\frac{a^{\prime \prime}(\eta)}{a(\eta)}\right) g_k = \frac{C_k(\eta)^2}{a^4(\eta)}g_k
\end{eqnarray}
and furthermore $C_k (\eta)$ would be determined by
\begin{equation}\label{eq48}
\frac{C_k (\eta)}{a^2(\eta)} = \frac{c_k}{f_k(\eta) g_k(\eta)},
\end{equation}
which we call $c_k$ the entanglement parameter and if this parameter vanishes then we regain (\ref{eq25}). Note that this parameter is constant of the integral and only depends on $k$. Here it is enough to solve coupled differential equations in (\ref{eq46}) with given initial conditions. ABH showed that these equations have a perturbative solution of even orders of entanglement parameter $c_k^{2n}$ for $f_k$ and odd orders $c_k^{2n+1}$ for $C_k$. In fact if $v_k$ and $u_k$ be solutions of eq. (\ref{eq47}) for $f_k$ and $g_k$ respectively when $c_k = 0$, then they found 
\begin{equation}\label{eqn48}
f_k (\eta) = v_{k} (\eta) \left( 1+ c_k^2 \mathcal{F}_k (\eta) + \cdots \right) ,
\end{equation}
\begin{equation}\label{eqn49}
g_k (\eta) = u_{k} (\eta) \left( 1+ c_k^2 \mathcal{G}_k (\eta) + \cdots \right) 
\end{equation}
and 
\begin{equation}\label{eqn50}
C_{k} = \frac{c_k}{u_k (\eta) v_{k}(\eta)} - \frac{c_k^3}{u_k (\eta) v_{k}(\eta)} \left( \mathcal{F}_k (\eta) + \mathcal{G}_k (\eta) \right) + \cdots ,
\end{equation}
where 
\begin{equation}\label{eqn51}
\mathcal{F}_k (\eta) =  \left( \int_{\eta_0}^{\eta} d\eta_1 \frac{1}{v_{k}^2 (\eta_{1})} \int_{\eta_0}^{\eta_1} d\eta_2 \frac{1}{u_{k}^2 (\eta_{2})}  \right) ,\ \ \ \ \ \ \ \ \ \ \mathcal{G}_k (\eta) =  \left( \int_{\eta_0}^{\eta} d\eta_1 \frac{1}{u_{k}^2 (\eta_{1})} \int_{\eta_0}^{\eta_1} d\eta_2 \frac{1}{v_{k}^2 (\eta_{2})}  \right).
\end{equation}
There some questions might arise about the entity of this state. We know that this state can not be a vacuum state and should be written in terms of a superposition of excited states of $\phi$ and $\chi$. As a result, to gain such view one has to go to Heisenberg language of the quantum field. In next section, we shall try to obtain this goal and find a solution for \eqref{eq46}. Note that the eq. (\ref{eq44}) is not the general entanglement solution for the system under study because it can not include all entangled states such as a state of form $\vert 0_{\phi} , 1_{\chi}\rangle + \vert 1_{\phi} , 0_{\chi}\rangle$. \\

\section{Entanglement Transformation}

In this section, we introduce a transformation like the Bogoliubov transformation which help us to reconstruct the previous entangled state in the Heisenberg picture. For simplicity, consider annihilation operator (\ref{eq27}) for $\phi$ in the new form 
\begin{equation}\label{eq49}
a_k = \mu_k \phi_{k} (\eta) +i\sigma_k  \frac{\delta}{\delta \phi_{-k}}  
\end{equation}
where
\begin{equation}\label{50}
\mu_k \equiv \frac{1}{i\sqrt{2}} \left[(a v_{k}^{'} (\eta)-a^{'} v_{k} (\eta)) \right]
\end{equation}
and
\begin{equation}\label{eq51}
\sigma_k \equiv \frac{1}{i\sqrt{2}} \frac{v_{k}}{a}(\eta).
\end{equation}
The similar annihilation operator for $\chi$
\begin{equation}\label{eq52}
b_k = \tilde{\mu}_k \chi_{k} (\eta) +i\tilde{\sigma}_k  \frac{\delta}{\delta \chi_{-k}}  ,
\end{equation}
where $\tilde{\mu}$ and $\tilde{\sigma}$ are similar to ${\mu}$ and ${\sigma}$ just with this different that $v_k \rightarrow u_k$ and $u_k \rightarrow v_k$. Note that we have absorbed $(2\pi)^3$ factor just by redefinition of commutators (\ref{eq20}) and (\ref{eq22}). Now, we transform ${a}_{k}$ and ${b}_{k}$ to the new annihilation and creation operators as following
\begin{equation}\label{eq53}
\tilde{a}_{k} = {\alpha}_k (\eta) a_{k} + {\beta}^{*}(\eta)_{k} b_{-k}^{\dagger}
\end{equation}

\begin{equation}\label{eq54}
\tilde{b}_{k} = {\alpha}_k (\eta) b_{k} + {\beta}^{*}_{k} (\eta) a_{-k}^{\dagger}.
\end{equation}
Here $\alpha_k$ and $\beta_k$ are like what we saw in Bogoliubov transformation with this difference that a Bogoliubov transformation is time-independent and so doesn't change the space of mode functions. This transformation just transforms a set of two independent mode functions into another set, but above entangled transformation mixes modes of different fields and also can have time-dependency. Unfortunately, Kanno \citep{Kanno:2015ewa} supposed that this kind of transformation as a time-independent transformation without any reason. Here we consider the general case and suppose this transformation can be time-dependent and finally describe why this can not be independent of time.
If we request that these new operators satisfy their own commutation relations like (\ref{eq22}) then we get
\begin{equation}\label{eq55}
|{\alpha}_k (\eta)|^2  - |{\beta}_{k} (\eta)|^2 = 1.
\end{equation}
Consequently, one can find out that the new vacuum state $\vert \tilde{0}\rangle$ (which annihilated by $\tilde{a}_{k}$ and $\tilde{b}_{k}$) is
 \begin{equation}\label{eq56}
\vert \tilde{0}\rangle = N(\eta)\exp \left( -\sum_k \frac{\lambda_k (\eta)}{2} a_{k}^{\dagger} a^{\dagger}_{-k} \right) \ \vert 0 \rangle ,
\end{equation}
where $\lambda_k = \frac{\beta^{*}_k}{\alpha_k}$. Now, let us find these states in Schr\"{o}dinger picture. To this end, we consider the following relations
\begin{equation}\label{eq57}
\langle \phi_{k} , \chi_{k} \vert \tilde{a}_{k} \vert \tilde{0} \rangle =\langle \phi_{k} , \chi_{k} \vert \tilde{b}_{k} \vert \tilde{0} \rangle = 0
\end{equation}
and use (\ref{eq49}) and (\ref{eq52}) to obtain
\begin{equation}\label{eq58}
{\alpha}_k \left[ \mu_k \phi_{k} (\eta) \Psi[\phi, \chi] +i\sigma_k  \frac{\delta \Psi[\phi, \chi]}{\delta \phi_{-k}} \right] + {\beta}^{*}_{k} \left[ \tilde{\mu}^{*}_k \chi_{k} (\eta)\Psi[\phi, \chi] +i\tilde{\sigma}^{*}_k  \frac{\delta\Psi[\phi, \chi]}{\delta \chi_{-k}} \right] = 0
\end{equation}
and
\begin{equation}\label{eq59}
{\alpha}_k \left[ \tilde{\mu}_k \chi_{k} (\eta) \Psi[\phi, \chi] +i\tilde{\sigma}_k  \frac{\delta \Psi[\phi, \chi]}{\delta \chi_{-k}} \right] + {\beta}^{*}_{k} \left[ {\mu}^{*}_k \phi_{k} (\eta)\Psi[\phi, \chi] +i{\sigma}^{*}_k  \frac{\delta\Psi[\phi, \chi]}{\delta \phi_{-k}} \right] = 0.
\end{equation}
These are just two simple functional differential equations which can be  solved directly. The interesting point is that the solution has the form of (\ref{eq44}) with the following kernels
\begin{equation}\label{eq60}
A_k = -i a^2 \left( \frac{v_{k}^{'}}{v_k} - \frac{a^{'}}{a} \right) + \lambda_k C_k \frac{u^{*}_{k}}{v_k} 
\end{equation}
and
\begin{equation}\label{eq61}
A_k = -i a^2 \left( (\frac{v_{k}^{'}}{v_k})^{*} - \frac{a^{'}}{a} \right) + {\lambda}_{k}^{-1} C_k \frac{u_{k}}{v_{k}^{*}} 
\end{equation}
and the same for $B_k$ would be obtain only by  $v_k \rightarrow u_k$ and $u_k \rightarrow v_k$. These are two algebraic equations which have the following solutions

\begin{equation}\label{eq62}
C_k (\eta) = \frac{2 a^2 (\eta)}{-\lambda_k (\eta) u_{k}^{*} (\eta) v_{k}^{*} (\eta) + {\lambda}_{k}^{-1} (\eta) u_{k} (\eta) v_{k} (\eta)}
\end{equation}
and
\begin{equation}\label{eq63}
A_k (\eta) = -i a^2 \left( \frac{v_{k}^{'}}{v_k} (\eta) - \frac{a^{'}}{a} (\eta) \right) + \lambda_k (\eta) \frac{2 a^2 (\eta)}{-\lambda_k (\eta) u_{k}^{*}(\eta) v_{k}^{*}(\eta) + {\lambda}_{k}^{-1}(\eta) u_{k}(\eta) v_{k}(\eta)} \frac{u^{*}_{k}}{v_k}(\eta).
\end{equation}
If the reader looks at the Kanno's paper, he or she will find out that the Kanno's results for $A_k, B_k$ and $C_k$ is different from what we have derived here. This happens because Kanno did't do the field redefinition and also wrote a wrong condition for the wronskian of  mode functions. 
Now, we request that the new vacuum to be a solution of Schr\"{o}dinger equation and with this consideration that we should equate these kernels with those we have found in (\ref{eq45}) and (\ref{eq46}). This helps us to find expressions for $f_k$ and $g_k$ which defined in (\ref{eq47}) as following
\begin{equation}\label{eqn1}
{f_k} = {\rho _k}{v_k}(\eta )\left[ {\exp \left( { 2i\int_{{\eta _0}}^\eta  {\frac{{{u_k}^*}{\lambda _k}}{{{-\lambda _k}{u^*}_k|{v_k}{|^2} + \lambda _k^{ - 1}{u_k}v_k^2}}} } \right)} \right]
\end{equation}
and
\begin{equation}\label{eqn2}
{g_k} = {\zeta _k}{u_k}(\eta )\left[ {\exp \left( { 2i\int_{{\eta _0}}^\eta  {\frac{{{v_k}^*}{\lambda _k}}{{{-\lambda _k}{v^*}_k|{u_k}{|^2} + \lambda _k^{ - 1}{v_k}u_k^2}}} } \right)} \right]
\end{equation}
where ${\rho _k}$ and $ {\zeta_k}$ are constants. We can see the way which we passed to finding an obvious description of the entangled state (by using of entanglement transformation), leads us to the above relations for non-linear equations in (\ref{eq47}). In the above relations to do the integrals in the exponents we, have to know the time-dependent form of $\lambda_k$. But what does happen if one supposes that $\lambda_k$ is time-independent? Let us to look at it more precisely. Comparing eq. (\ref{eq62}) with eq. (\ref{eqn50}) we find out that $\lambda_k$ should be a function of entanglement parameter $c_k$. In fact $c_k = 0$ implies that $\lambda_k = 0$. When $c_k \ll 1$ one can show that $\lambda_k = c_k /2$. As a result, $\lambda_k$ is approximately constant and eqs. (\ref{eqn1}, \ref{eqn2}) are consistent with (\ref{eqn48}, \ref{eqn49}). However when we go ahead  and consider the next powers of $c_k$ we find out that  $\lambda_k$ can not be time-independent. If we expand $\lambda_k$ in terms of the entanglement parameter, one can show that
\begin{equation}
\lambda_k (\eta) = c_k /2 - c^3_k  N_k (\eta) + \cdots,
\end{equation}
where 
\begin{equation}
N_{k}(\eta) = \frac{\mathcal{F}_k + \mathcal{G}_k}{2 a^2 } (\eta) + \frac{1}{8} \frac{u_k^{*} v_k^{*}}{u_k v_k} (\eta)
\end{equation}

This implies that the entangled state defined by entanglement transformation will coincide with the ABH state if  $\lambda_k$ is time-dependent.
It is interesting to calculate the power spectrum in this picture. Kanno \cite{Kanno:2015ewa} has shown that the power spectrum of $\phi$ in Schr\"{o}dinger picture is the same as the Heisenberg picture power spectrum only up to the second order of $\beta_k$. Kanno expands the Albertch et.al's result (for power spectrum in the Schr\"{o}dinger picture) with smallness assumption of $\beta_k$. She didn't get to her exact form for power spectrum which she finds in the Heisenberg picture  \cite{Kanno:2015ewa}. In contrast with Kanno's analysis, we find out in our analysis the two power spectra do coincide exactly without any approximation. In fact, with make using of (\ref{eq62}) and (\ref{eq63}) and after some tedious calculations one can show that
\begin{equation}
\langle\tilde{0}\vert\phi_{-k}\phi_k \vert \tilde{0} \rangle = \frac{1}{2 a^2}|v_k (\eta)|^{2} \left( \frac{1+ |\lambda_k|^2}{1- |\lambda_k|^2}\right) = \frac{1}{2} \frac{B_{kR}}{A_{kR} B_{kR} - C^{2}_{kR}},
\end{equation}
in which the sub-index $R$ stands for the real part of each variable. One could see in this analysis,  we are not only able to solve mode functions of entangled space in term of entanglement transformation parameter but also re-derive the power spectrum which people have found in Schr\"{o}dinger analysis. There just remains a point which we should clarify and that is by using Heisenberg picture, one can access other entangled states in a simple way. It is easy to see from (\ref{eq56}) that the ABH state creates an entanglement between states of two fields with the same number of particles i.e. $\vert 0_{\phi} , 0_{\chi}\rangle$ , $\vert 1_{\phi} , 1_{\chi}\rangle$, \ldots . To make an entangled state in which states with different fields with a different number of particles are entangled with each other, it is better (and usual) to go the Heisenberg picture and by using of creation operators create an excited state of arbitrary number of particles and then mix the resultant state with momentum dependent coefficients. Consequently, without concern, the final entangled state is a solution for Schr\"{o}dinger equation.\\

\section{Conclusion}

An important aspect of the quantum physics is the entanglement which can have the observational effect on cosmic structure formation which has originated from quantum fluctuations in the early universe \cite{En-Observational}. To study the entanglement in the early universe, first Albrecht e.t \cite{Holman-entanglement} have shown that the initial entangled state can have some features on the observable power spectrum in the early universe and then Kanno attempted to construct these initial entangled state by making use of a transformation which was similar to the Bogoliubov transformation \cite{Kanno:2015ewa}. Besides showing that the Kanno's calculation and analysis were not correct in some parts, in this article we have explored an entangled transformation (which is introduced Heisenberg picture) role on the  Schr\"{o}dinger field theory of the FLRW model. In accordance with this article analysis, we conclude with some remarks:

\begin{itemize}
   \item The solution of the entangled coefficient $A, B, C$ in the  Schr\"{o}dinger picture can be written in terms of the time-dependent entangled transformation parameter in the Heisenberg formalism. The entangled coefficient in (\ref{eq53}) is not the Bogoliubov transformation since it mixes the mode functions of different fields and can not be interpreted as a different observer physics like in the Bogoliubov transformation.
   
   \item   We should mention that an entangled state (see Eq.(\ref{eq44})) defined by an Entanglement transformation, is not in general a solution of Schr\"{o}dinger equation and one has to use Schr\"{o}dinger equation to find parameter$ \lambda_k$ of the entanglement transformation while a Bogoliubov transformation keeps us in space of solutions of Schr\"{o}dinger equation.

    \item  We calculated the relation between two vacuum states which their mode functions are transformed by the Bogoliubov transformation in the  Schr\"{o}dinger picture.

    \item We have derived the entangled parameter $C_k$ in the  Schr\"{o}dinger picture in term of the entangled transformation parameter $\lambda_k$ in the Heisenberg picture. This formalism verifies that the entanglement transformation depends on the conformal time.
    
    \item The power spectrum of the inflaton field can be presented both in terms of the entangled parameter of  Schr\"{o}dinger picture and the entangled transformation coefficient in the Heisenberg one which has the same value.
    
    \item It was discussed and shown that the entanglement has a more deep meaning which can not be presented in the form of the (\ref{eq44}) and in term of the entanglement transformation (\ref{eq53}).
     We have presented an example that cannot describe the entanglement  (\ref{eq44}) and (\ref{eq53}). 
    Moreover, the entanglement was discussed in the equation (\ref{eq44}) and (\ref{eq53}) describe the entanglement between two fields and do not quantify the entanglement between the excited state of each field.\\

\end{itemize}




\end{document}